# Genetic structure of Sino-Tibetan populations revealed by forensic STR loci

Hong-Bing Yao[1,†], Chuan-Chao Wang[2,†], Jiang Wang[3], Xiaolan Tao[1], Shao-Qing Wen[2], Qiajun Du[4], Qiongying Deng[5], Bingying Xu[6], Ying Huang[6], Hong-Dan Wang[7], Shujin Li[8], Bin Cong[8], Liying Ma[1], Li Jin[2,9], Johannes Krause[10], Hui Li[2,]*

1. Key Laboratory of Evidence Science of Gansu Province, Gansu Institute of Political Science and Law, Lanzhou 730070, China
2. State Key Laboratory of Genetic Engineering and Ministry of Education Key Laboratory of Contemporary Anthropology, Collaborative Innovation Center for Genetics and Development, School of Life Sciences, Fudan University, Shanghai, 200433, China
3. College of Animal Sciences and Veterinary Medicine, Henan Agricultural University, Zhengzhou 450002, Henan Province, China
4. Lanzhou University Second Hospital Clinical Laboratory, Lanzhou 730000, Gansu Province, China
5. Department of Anatomy, Guangxi Medical University, Nanning 530021, China
6. School of Forensic Medicine, Kunming Medical University, Kunming, 650500, China
7. Medical Genetic Institute of Henan Province, Henan Provincial People's Hospital, People's Hospital of Zhengzhou University, Zhengzhou, China
8. Hebei Key Laboratory of Forensic Medicine, Department of Forensic Medicine, Hebei Medical University, Shijiazhuang, 050017, China
9. CAS-MPG Partner Institute for Computational Biology, Shanghai Institutes for Biological Sciences, Chinese Academy of Sciences, 200031 Shanghai, China
10. Max Planck Institute for the Science of Human History, Kahlaische Straße 10, 07745 Jena, Germany

†These authors contributed equally to this work.
*Corresponding author: Prof. Hui Li. Tel: +86-21-51630427, E-mail addresses: LHCA@fudan.edu.cn

**ABSTRACT**

The origin and diversification of Sino-Tibetan populations have been a long-standing hot debate. However, the limited genetic information of Tibetan populations keeps this topic far from clear. In the present study, we genotyped 15 forensic autosomal STRs from 803 unrelated Tibetan individuals from Gansu Province (635 from Gannan and 168 from Tianzhu). We combined these data with published dataset to infer a detailed population affinities and admixture of Sino-Tibetan populations. Our results revealed that the genetic structure of Sino-Tibetan populations was strongly correlated with linguistic affiliations. Although the among-population variances are relatively small, the genetic components for Tibetan, Lolo-Burmese, and Han Chinese were quite distinctive, especially for the Deng, Nu, and Derung of Lolo-Burmese. Southern indigenous populations, such as Tai-Kadai and Hmong-Mien populations, might have made substantial genetic contribution to Han Chinese and Altaic populations, but not to Tibetans. Likewise, Han Chinese but not Tibetan shared very similar genetic makeups with Altaic populations, which did not support the "North Asian origin" of Tibetan populations. The dataset generated here are also valuable for forensic identification.

**INTRODUCTION**

The Sino-Tibetan languages, spoken by over a billion people all over East Asia and Southeast Asia, have been classified into two subfamilies, namely Chinese and Tibeto-Burman[1]. The linguistic connection between Chinese and Tibeto-Burman are well established. Chinese was suggested to split away from Tibeto-Burman around 6 thousand years ago (kya) based on lexical evidence[2].

During the past two decades, genetic evidence, especially from the maternal mitochondrial DNA (mtDNA) and the paternal Y chromosome, has shed more light on the history of Sino-Tibetan populations. MtDNA evidence revealed a northern Asian origin of Tibetans, due to the high frequencies of northern Asian specific haplogroup A, D, G, and M8[3-5]. The genetic relics of the Late Paleolithic ancestors of Tibeto-Burman populations have also been reported, such as haplogroup M62[5]. Y chromosome suggested Tibeto-Burman populations are an admixture of the northward migrations of East Asian initial settlers with haplogroup D-M175 in the Late Paleolithic age, and the southward Di-Qiang people with dominant haplogroup O3a2c1*-M134 and O3a2c1a-M117 in the Neolithic Age[6]. Haplogroup O3a2c1*-M134 and O3a2c1a-M117 are also characteristic lineages of Han Chinese, comprising 11.4% and 16.3%, respectively[7,8]. However, another dominant paternal lineage of Han Chinese, haplogroup O3a1c-002611, was found at very low frequencies in Tibeto-Burman populations, suggesting this lineage might not have participated in the formation of Tibeto-Burman populations[6, 9]. Sex-biased admixture has also been observed during the formation of Tibeto-Burman populations. Southern Tibeto-Burman populations exhibited a stronger influence of northern immigrants on the paternal lineages and a more extensive contribution of southern natives to the maternal lineages[10]. Likewise, the southern natives made a greater contribution to the maternal lineages of southern Han Chinese[11]. A genome-wide study of PanAsia SNP project revealed that Han Chinese populations show varying degrees of admixture between a northern Altaic cluster and a Sino-Tibetan/Tai-Kadai cluster[12]. But Tibetan populations were not included in the PanAsia project. Tibeto-Burman populations tended to cluster with North Asian (Altaic) and Tai-Kadai populations rather than Han Chinese based on the frequency data of 15 autosomal short tandem repeats (STRs)[13].

From previous studies, the origin of Sino-Tibetan populations seems to involve substantial genetic admixture with surrounding populations. However, the limited markers of mtDNA and Y chromosome and small sample sizes of genome-wide study are far from enough to give a comprehensive understanding about the genetic history and admixture process of Sino-Tibetan populations. In addition, Tibetan populations of Gansu province, the key area for the diversification of Amdo Tibetan, have seldom been studied genetically. Therefore, we analyzed 15 autosomal STRs in 635 and 168 unrelated individuals from two Tibetan populations in Gannan and Tianzhu of Gansu province to explore the genetic structure of Tibetan populations and to test population affinities and the level of admixture of Sino-Tibetan populations with surrounding populations.

**MATERIALS AND METHODS**

We collected blood samples of 635 and 168 unrelated individuals from two Tibetan populations in Gannan and Tianzhu, Gansu province. Our study was approved by the Ethnic Committee of Gansu Institute of Political Science and Law. All individuals were adequately informed and signed their informed content before their participation. For each sample, genomic DNA was extracted according to the Chelex-100 method and proteinase K protocol[14]. 15 most widely used forensic loci were amplified simultaneously using AmpFlSTR Sinofiler PCR Amplification Kit (Applied Biosystems, Foster City, CA, USA) at the D8S1179, D21S11, D7S820, CSF1PO, D3S1358, D13S317, D16S539, D2S1338, D19S433, vWA, D18S51, D5S818, FGA, D6S1043 and D12S391 STR loci. The PCR products were analyzed with the 3500XL DNA Genetic Analyzer and Genemapper ID-X software (Applied Biosystems, Foster City, CA, USA).

Allele frequency, heterozygosity, polymorphism information content (PIC), discrimination power (DP), probability of paternity exclusion (PPE) were calculated using PowerStatesV12 (http://www.promega.com/). Tests for Hardy–Weinberg equilibrium were performed in Arlequin v3.5.1.3[15]. Since the statistical analyses in this study were on the basis of Bayesian-clustering algorithm, raw genotypic data of 13 STRs (excluding D6S1043 and D12S391) from 59 populations all around the world were extracted to determine population affinity (Dongxinag and Hui of Gansu are unpublished data of our lab)[13,16-44]. Analysis of molecular variance (AMOVA), average number of pairwise differences, pairwise Fst, Slatkins linearized Fst, and coancestry coefficients were all calculated in Arlequin v3.5.1.3[15] using genotype data. The detailed population genetic structure was performed using model-based clustering method implemented in Structure 2.3.4[45,46] under assumptions of admixture, LOCPRIOR model, and correlated allele frequencies. Each run used 100,000 estimation iterations for K = 2 to 12 after a 20,000 burn-in length with several replicates. Posterior probabilities for each K were computed for each set of runs. Graphical display for Matrix plot of genetic distance and population structure were carried out in R statistical software v3.0.2[47] and Distruct v1.1[48].

**RESULTS**

Fifteen STR loci were typed in two populations sampled from Gannan and Tianzhu of Gansu province and their allele frequencies along with a number of genetic and forensic parameters of interest are provided in Supplementary Table 1 and 2. No significant deviation was observed for Hardy–Weinberg equilibrium tests, indicating that our samples well represent the populations. The loci in both populations were highly discriminating with DP ranging from 0.852 to 0.974, demonstrating that this set of loci are useful for forensic identification.

We performed various parameters of genetic diversity and distances to infer population structure between Tibetans in Gannan and Tianzhu, as well as compared them with previously studied populations. The within-population component of genetic variation, estimated here as 99.14% (Table 1), accounts for most of genetic diversity of the 20 Sino-Tibetan populations. The small among-population and among-group variance components supported the genetic affinity among the Sino-Tibetan populations. The pairwise Fst comparison also confirmed the genetic similarity between Tibetan and Han Chinese populations, with almost all the values below 0.01. However, Deng, Nu, and Derung seemed like to be outliers of the Sino-Tibetan profile due to almost all the Fst values between each of them with other Sino-Tibetan populations are above 0.03. The Fst between other Lolo-Burmese populations with Tibetan or Han Chinese were also slightly higher than the values between Tibetan and Han Chinese. Tibetan and Han Chinese also showed close genetic relationship with Altaic population Russian (Inner Mongolia) and Korean in East Asia, but not with those in South Siberia, such as Buryat, Altay, Tofalar, Sojot, and Khakas. The genetic distances between Tibetan and Han Chinese with Tai-Kadai (Maonan, Mulam, and Thai) or Hmong-Mien (She) populations were also larger than those between Tibetan and Han Chinese. The Muslim populations in northwest China exhibited relatively small genetic distances with Tibetan and Han Chinese, revealing the substantial gene flow from Sino-Tibetan populations into Muslim people during their Islamization. Average number of pairwise differences, Slatkins linearized Fst, and coancestry coefficients also revealed a very similar pattern with the pairwise Fst (Supplementary Table 3).

We then applied a model based clustering algorithm in Structure to infer the detailed genetic ancestry at individual level. This approach will place individuals into K clusters, where K is set in advance but can be varied. The results for K=2 to 7 are shown in Figure 2 and Supplementary Table 4. At K=2, a clear distinction was observed between the European and African with populations from Asia. At K=3, Lolo-Burmese populations, especially Deng, Nu, and Derung, were separated from Asian populations. At K=4, Tibetan populations were identified from other Asian populations and this genetic component also had proportions of 20%-50% in Han Chinese and Altaic populations, but greatly reduced in Maonan. At K=5, European was separated from African. Uygur and Altaic populations in southern Siberia seemed to have half of this European component. The next cluster at K=6 corresponded to Tai-Kadai and Hmong-Mien populations in south China. This southern native component comprised almost half of the Han Chinese and Altaic gene pool. At K=7, the southern native genetic component in Han Chinese and Altaic populations formed a new cluster. It's quite clearly from the Structure analysis that Han Chinese and Altaic populations of East Asia share the similar membership, but the Altaic population in southern Siberia shared more genetic component with European. The origin of Tibetan populations seemed to involve gene flow from Han Chinese and Altaic populations. Southern natives, such as Tai-Kadai and Hmong-Mien, likely made substantial genetic contribution to Han Chinese, Altaic populations, and Yi in Yunnan, but not to Tibetan populations. In addition, Tibetan in Tianzhu of Gansu province seemed to have more Han Chinese and Altaic genetic components than other Tibetan populations had.

## DISCUSSION
The origin and diversification of Sino-Tibetan populations have been a long-standing hot topic among linguists, population geneticists, anthropologists, and archaeologists. However, the limited genetic information of Tibetan populations has made this topic far from clear. Here, we typed 15 forensic autosomal STRs from 635 and 168 unrelated Tibetan individuals from Gannan and Tianzhu of Gansu province, together with published forensic dataset to infer a detailed genetic structure of Sino-Tibetan populations. Tibetan of Gannan shared a very similar genetic makeup with other Tibetan populations from Tibet, Qinghai, and Yunnan. While Tibetan of Tianzhu County seemed like to share more genetic component with Han Chinese and Altaic populations, which is understandable as those Tibetans are surrounded by Chinese Muslims and Mongolic people. The genetic structure of studied Sino-Tibetan populations is strongly correlated with linguistic affiliations, as we can detect three distinctive genetic components for Tibetan, Lolo-Burmese, and Han Chinese although the among-population variances are relatively small. Yi of Yunnan province, one of Lolo-Burmese populations, was found out to be an admixture between Tibetan, Han Chinese, and southern natives (Tai-Kadai and Hmong-Mien). However, other Lolo-Burmese populations, such as Deng, Nu, and Derung, formed a distinctive cluster, which is probably due to isolation and genetic drift as those populations are all small and living on hunting and gathering[13].

Previous studies, especially using mtDNA and Y chromosome, had suggested the North Asian origin of Tibetan populations[49-50]. But this conclusion is ambiguous as the concept of "North Asian" was not clear and the results were mainly based on frequency data of two loci. Our results showed that the Tibetan were quite distinctive from Altaic populations. The genetic makeups of Altaic populations in East Asia are very similar with Han Chinese rather than with Tibetan populations. Furthermore, the Altaic population in southern Siberia, such as Buryat, Altay, Tofalar, Sojot, and Khakas, shared substantial genetic components with Europeans which are rarely seen in Sino-Tibetan populations. It's clearly that the formation of Tibetan populations surely involved gene flow from surrounding Han Chinese and Altaic people, but the words "North Asian origin" seem to exaggerate the influence. Further study using genome-wide markers will be necessary to reconstruct more authentic history of Tibetan populations.

**CONFLICT OF INTEREST**
The authors declare no conflict of interest.

**ACKNOWLEDGEMENTS**
This work was supported by the Natural Science Foundation of Gansu province (1308RJZA190), Scientific Research Project for Colleges of Gansu province (2014A-085), National Excellent Youth Science Foundation of China (31222030), National Natural Science Foundation of China (91131002), Shanghai Rising-Star Program (12QA1400300), MOE University Doctoral Research Supervisor's Funds (20120071110021), and MOE Scientific Research Project (113022A).

**Supplementary Material**
Supplementary Table 1. 15 autosomal STRs of Tibetan in Gannan and Tianzhu.
Supplementary Table 2. Allele frequency distributions and forensic parameters
Supplementary Table 3. Matrix of pairwise $F_{ST}$, Slatkins linearized $F_{ST}$, average number of pairwise differences, and coancestry coefficients.
Supplementary Table 4. Estimates of posterior probabilities of data under admixture model for 38 populations and the proportion of membership of each pre-defined population in each cluster.

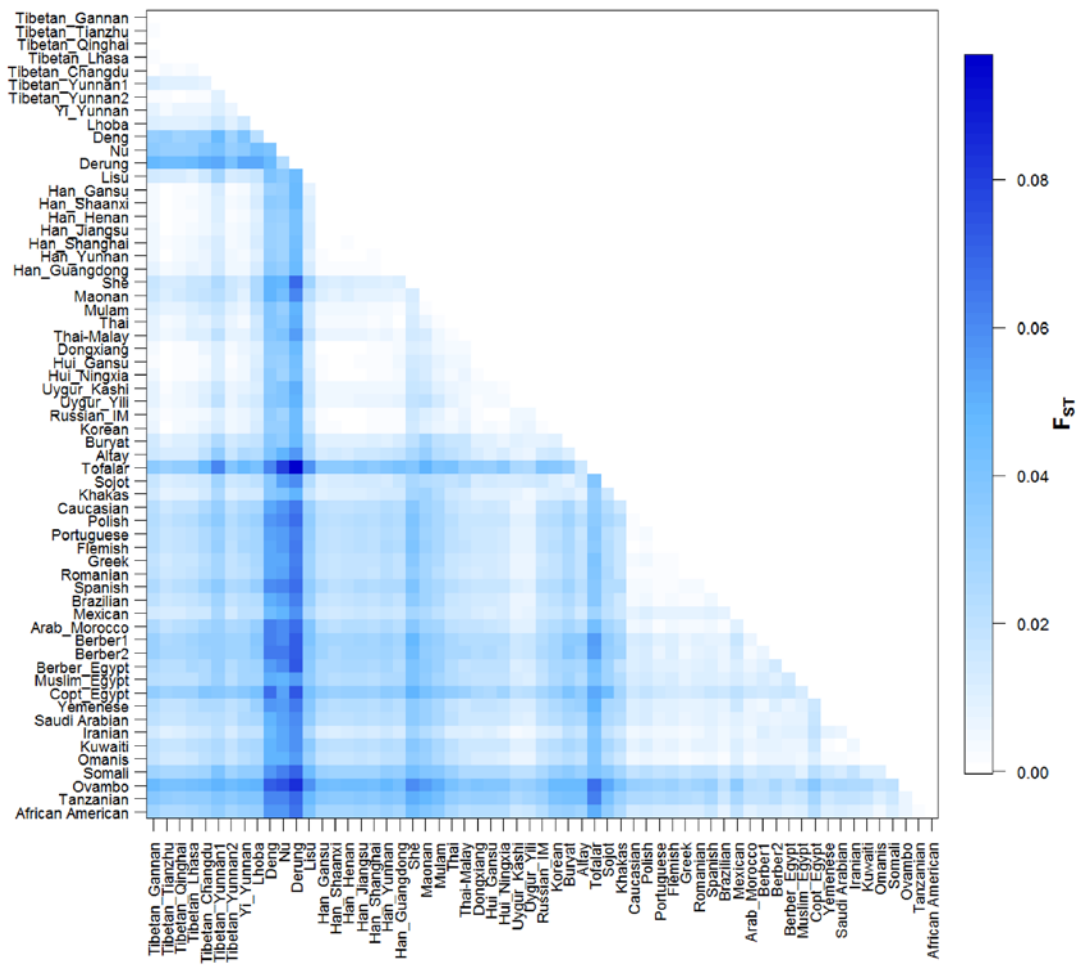

Figure 1. Plots of pairwise Fst of Tibetan in Gannan, Tianzhu and other 59 worldwide populations

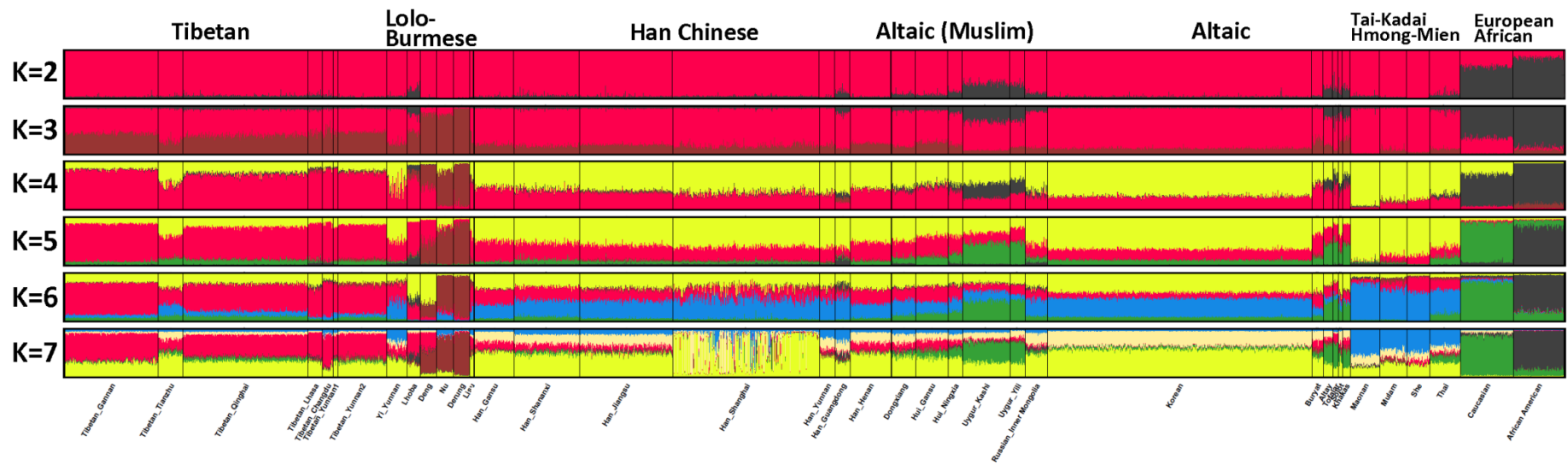

Figure 2. Estimated population genetic structure of Tibetan in Gannan and Tianzhu, and other 36 worldwide populations.

Table 1. AMOVA results for 13 autosomal STRs at population and group scales. 20 Sino-Tibetan populations have been classified into 3 groups: Han Chinese, Tibetan, and Lolo-Burmese.

| Source of variation | d.f. | Sum of squares | Variance components | Percentage of variation |
|---|---|---|---|---|
| Among group | 2 | 175.190 | 0.01925 | 0.37 |
| Among populations within groups | 17 | 306.945 | 0.02549 | 0.49 |
| Within populations | 11204 | 58131.8 | 5.18848 | 99.14 |
| Total | 11223 | 58613.9 | 5.23323 | |